\documentclass[11pt,letterpaper]{article}

\usepackage{authblk}
\usepackage{xcolor}
\usepackage[title]{appendix}

\title{Classical physics and human embodiment: \\
The role of contemplative practice in integrating formal theory and personal experience in the undergraduate physics curriculum}
\author[1]{Zosia Krusberg}
\author[2]{Meredith Jane Ward}
\affil[1]{Department of Physics and Astronomy, Northwestern University}
\affil[2]{Center for Anxiety and Traumatic Stress Disorders and Complicated Grief Program, Massachusetts General Hospital}
\date{}

\begin{document}
\maketitle

\begin{abstract}
One of the objectives of the undergraduate physics curriculum is for students to become aware of the connections between the fundamental principles of classical physics and their personal experience.  Nonetheless, numerous studies have shown that students' awareness of such connections tends to deteriorate, sometimes substantially, following instruction.  In this work, which constitutes the first analysis of the effects of contemplative practices on the learning experience in undergraduate physics courses, we present two practices aimed at integrating formal theory with students' personal, embodied experience: a sensory meditation and a contemplative videography.  In written reflections on their experiences with the practices, the students expressed suddenly becoming aware of countless manifestations of formal physics principles in their surroundings, in an important step toward establishing firm connections between the abstract and the experiential.  Furthermore, the students reported experiencing a sense of physical embodiment, heightened sensory awareness, somatic relaxation, and mental stillness, in significant contrast to their typical experience.  Students also described experiencing insight about the essential role of observation in the scientific endeavor, a reawakened sense of curiosity, an intrinsic motivation to understand the observed physical phenomena, and a deepened awareness of their cognitive and emotional processes.
\end{abstract}

\clearpage

\tableofcontents
\clearpage

\section{Introduction}

\subsection{The objectives of physics education}

A multitude of objectives, both grand and minute, shapes the undergraduate physics curriculum.  Most essentially, students should learn the fundamental concepts of classical, modern, and contemporary physics, sophisticated and broadly transferable critical thinking and problem solving skills, and experimental techniques of historical and contemporary significance.  Students should also develop an interdisciplinary perspective of the field of physics, including its conceptual and procedural connections to other science, technology, engineering, and mathematics (STEM) fields.  Finally, students should acquire an awareness of the applications of physics to the collective and individual human experience, such as its interplay with politics, religion, and the arts, and its power to make sense of our experiences as embodied beings in the physical universe \cite{Redish2003b,NRC2013}.  This study addresses the last of these objectives: the relationship between physics and the individual human experience.

\subsection{Physics and personal experience}

Fundamentally, physics consists of a collection of formal mathematical theories describing the physical phenomena accessible to our senses and their technological extensions.  It is thereby inherently grounded in the human experience.  More directly, however, our everyday experience abounds with manifestations of the very principles that compose the undergraduate physics curriculum: the fundamental principles of classical mechanics and electrodynamics, the laws of thermodynamics, and even the central tenets of special relativity and quantum mechanics.

Despite this inherent relationship between physics and personal experience, students' perceptions of such a relationship tends to deteriorate---sometimes substantially---following instruction.  Two surveys commonly used to measure students' beliefs about physics, the Maryland Physics Expectations Survey (MPEX) and the Colorado Learning Attitudes about Science Survey (CLASS), both contain clusters of questions that assess students' belief in a connection between their experiences inside and outside the physics classroom \cite{Redish1998,Adams2006}.  The MPEX, for instance, includes items such as ``Learning physics helps me understand situations in my everyday life'' and ``To understand physics, I sometimes think about my personal experiences and relate them to the topic being analyzed.''  Similarly, items in the CLASS include ``Learning physics changes my ideas about how the world works'' and ``I think about the physics I experience in everyday life.''  A recent meta-analysis of twenty-four studies using the MPEX and CLASS found that, in typical physics classes, the number of students agreeing with such statements decreased following instruction---in one extreme case, from 71\% to 52\%.  Physics majors were somewhat more likely than non-majors to agree with these statements prior to instruction; however, their beliefs did not improve significantly over their undergraduate careers \cite{Madsen2015}.  

Although this phenomenon has been described and discussed extensively in the educational literature, few reliable interventions have been designed and implemented to help students develop a strong connection between physics and personal experience.  Even courses with innovative teaching techniques that have successfully improved students' conceptual understanding and problem solving generally only attempt to help students develop such connections with example problems and homework problems that are nominally related to their experience \cite{Crouch2001,May2002,Wieman2005a,Chu2007,Crouch2014,Lasry2016,Webb2017}.  It is exceedingly rare that students' personal, embodied experience is invoked and discussed---or even acknowledged---in the contemporary physics classroom.

\subsection{Contemplative practice in higher education}


Contemplative practices have long held an essential place in many of the world's spiritual and philosophical traditions.  Such practices manifest in countless forms: simple moments of silence, sitting meditation, embodied movement, deep listening, and practices cultivating compassion and loving-kindness. Despite this apparent variety, all contemplative practices share an introspective focus that fosters an awareness of one's experience in the present moment \cite{Barbezat2014}.  

In recent years, there has been a surge of interest in the application of contemplative practices to secular settings in forms that do not require adherence to specific cultural or religious beliefs.  In medicine and clinical psychology, for instance, there is growing evidence of the positive impact of mindfulness-based stress reduction (MBSR) and mindfulness-based cognitive therapy (MBCT)---two treatment modalities based on mindfulness meditation---on patients suffering from anxiety and depression, as well as patients with a range of chronic stress-related and pain-related conditions \cite{Kabat-Zinn2006,Ospina2007,Bohlmeijer2010,Hayes2011,Kerr2013,Buchholz2015}.  Similarly, in educational institutions from preschool to graduate school, meditation practice has successfully improved the cognitive and emotional health of both students and teachers by improving focus and information processing; reducing stress, anxiety, and depression; increasing emotional regulation and the prevalence of positive emotional states; and developing creativity, empathy, and self-compassion \cite{Jha2007,Shapiro2008,Morrison2014,Rechtschaffen2014,Short2015,Palmer2017}.  

Meanwhile, a variety of contemplative practices are transforming teaching and learning in higher education, as instructors increasingly integrate them directly into course curricula.  By cultivating an awareness of the subjective experience, contemplative practices offer students an invitation to personal observation, exploration, and discovery of course material.  Furthermore, they constitute a potentially impactful supplement to traditional curricular activities---which typically emphasize rational and analytical thinking---allowing students to become aware of connections between the abstract and the experiential.  Over the last two decades, contemplative practices have been integrated into courses in the humanities, the social sciences, and the natural sciences, as well as in professional schools \cite{Hart2004,Zajonc2006a,Bush2011,Levy2016}.  In STEM education specifically, courses in mathematics, biology, environmental science, chemistry, earth science, and physics have explored the use of contemplative practices alongside more traditional assignments \cite{Barbezat2014,Schneiderman2013,Zajonc2013,Francl2016}.  To date, however, no analysis of the effects of contemplative practices on the learning experience in undergraduate physics courses has been carried out.

\subsection{Overview}

In this work, we present a qualitative, exploratory analysis of the impact of contemplative practices on the learning experience in the undergraduate physics curriculum.  In Section \ref{method}, we describe two contemplative practices---a sensory meditation and a contemplative videography---aimed at drawing students' attention to their embodied experience of physical phenomena described by the principles of classical physics.  Then, in Section \ref{results}, we report on common themes in the reflections on these practices from students at three different institutions of higher education, and present notable excerpts from these reflections.  Finally, in Section \ref{conclusion}, we offer some concluding thoughts.


\section{Method}
\label{method}

\subsection{The contemplative practices}

In the first practice, a sensory meditation inspired by the bodywork practices of Reginald Ray and David Rome, students are guided through a brief somatic meditation in which they direct their attention into and through their bodies, gradually expanding their awareness into the surrounding space \cite{Ray2008,Ray2016,Rome2014}.  They are then invited to use their senses to experience and explore the physical phenomena taking place around them.  

In the second practice, a contemplative videography inspired by Ch\"{o}gyam Trungpa Rinpoche's teachings on Dharma Art and the work of contemplative photography teachers Andy Karr and Michael Wood, students are asked to select one concept in classical physics on which to focus their attention for a full week \cite{Trungpa2008,Karr2011}.  They are then instructed to capture manifestations of their chosen concept in their surroundings on video, and combine their footage into short, focused films.  

The written guidelines for both practices can be found in Appendix A and B.

\subsection{Setting}

The two contemplative practices were integrated into existing curricula in five calculus-based courses in classical mechanics and electrodynamics, aimed primarily at students majoring in the physical sciences, engineering, and mathematics.  One course took place at a private, independent, coeducational liberal arts college, and the other four at private research universities.  In each course, students were required to complete at least one of three contemplative practices; the third practice was a reflective writing assignment not discussed in this work.  All courses were taught by the first author.  Student participation in the contemplative practices is outlined in Table 1.

\begin{table}[ht]
\begin{center}
\begin{tabular}{| l | c | c |}
\hline
\hline
Course & Sensory meditation & Contemplative videography \\
\hline
1 & 20 & -- \\
2 & 71 & 26 \\
3 & 71 & 38 \\
4 & 16 & 12 \\
5 & 12 & 5 \\
\hline
\hline
{\bf Total:} & {\bf 190} & {\bf 81} \\
\hline
\hline
\end{tabular}
\end{center}
\caption{Student participation in the contemplative practices}
\end{table}

\clearpage

\subsection{Written reflections}

To assess the impact of the two contemplative practices, students were required to submit written reflections on each practice.  The reflection prompts were deliberately vague, requesting simply that students discuss their experience completing the practices and whether they felt that contemplative practices could complement the formal study of physics.  In order to encourage students to be genuine in their assessment of the practices, the reflections were only graded on completion.  The reflections were collected electronically, catalogued without personally identifying information, and, finally, read and coded.  The coding procedure involved two readings of the corpus: in the first reading, emergent themes were identified, and, in the second reading, phrases and paragraphs were highlighted and labeled according to association with a given theme.  In order to ensure consistency, all coding was completed by one of the authors.

\section{Results}
\label{results}

\subsection{Summary of results}

A summary of common themes in the reflections on the sensory meditation and contemplative videography, along with their occurrence in the corpus, is shown in Tables 2 and 3.  In the subsequent sections, we discuss each of the themes---with extracts from the students' reflections---in more detail.

\begin{table}[b]
\begin{center}
\begin{tabular}{| l | c |}
\hline
\hline
Theme & Occurrence \\
\hline
Sense of physical embodiment and sensory awareness & 97\% \\
Awareness of applicability of formal theory to personal experience & 77\% \\
Awareness of cognitive and emotional processes & 76\% \\
Feeling of curiosity in response to observations & 72\% \\
Feeling of somatic relaxation and mental stillness & 66\% \\
Feeling of intrinsic motivation to study physics & 49\% \\
Recognition of role of observation in science & 37\% \\
\hline
\hline
\end{tabular}
\end{center}
\caption{Sensory meditation themes ($N = 190$)}
\end{table}

\begin{table}[h]
\begin{center}
\begin{tabular}{| l | c |}
\hline
\hline
Theme & Occurrence \\
\hline
Awareness of applicability of formal theory to personal experience & 99\% \\
Feeling of intrinsic motivation to study physics & 52\% \\
Feeling of curiosity in response to observations & 40\% \\
Awareness of cognitive and emotional processes & 30\% \\
\hline
\hline
\end{tabular}
\end{center}
\caption{Contemplative videography themes ($N = 81$)}
\end{table}

\subsection{Applicability of formal theory to personal experience}

In their written reflections, nearly all students expressed that carrying out the contemplative practices made them highly aware of the applicability of formal physics principles to their observations and embodied experience.

\begin{quote}
I see people strolling quietly through the courtyard. And I start thinking, if it were not for gravity, these people would be floating around, they would not be bound to Earth's surface. Without gravity, life as we know it would not be a possibility; we would not be able to walk or jump. Gravity is just something we all take for granted. Wow, Newton! In one of the rooms upstairs, someone is having an intense conversation on the phone. I can hear it too, quite clearly. I think about sound, about sound waves. Sound, a wave that is carried through air and other materials. Music starts playing in my head. Oh my God, without sound there would be no music; even worse, there would no verbal communication between people! I'm sitting on this chair. I take that in for a second. I think about the electrical repulsion between the electrons in the chair (or floor) and the electrons in my body---that frictional force that keeps me from falling through the chair. Also Newton's third law in application here, since I'm not squashing down the chair. Friction. The reason my clothes do not slip off my body; the reason my hands warm up when I rub them against each other in the cold; the reason why I can pull the brakes to bring my car to a standstill. I inhale, and smell the blooming flowers, I can smell spring in the air.  
\end{quote}

\begin{quote}
Physics and the arts are simply different languages used to describe the same phenomena. Lying in the hammock, I might have been thinking about Chopin, but at one point, I even started thinking about how everything I was seeing could in fact be explained by ideas we learned in class. The rocking hammock was just like the simple harmonic oscillators we played around with in physics lab. The sights of the beautiful blue sky and the vibrant green grass were dependent on principles of optics. The chirping of the birds was being carried to me by sound waves. It occurred to me that physicists and artists share the same source of inspiration---a world that is complicated and interesting and that makes us wonder about it and wonder about ourselves.  
\end{quote}

\begin{quote}
Looking again at the crashing of the waves into the rocks reminded me of the animations of waves interacting with soft boundaries  since nothing physically pins down the edge of the wave, the wave is able to run up the side of the rock upon contact, and the rock acts as a soft boundary to allow temporary constructive interference within that wave, before sending the wave back the opposite direction. It was an insane realization for me: the waves before me literally mirrored the animation from class!  
\end{quote}


\begin{quote}
When lying back and taking in the stars I find it pretty easy to get caught up in all the things between me and them: the distance, the air molecules buzzing and swarming around, probably colliding and interacting with the light before it even gets to me. Arguably they're blocking the view, but I suppose there's an equal argument to be made that they {\it are} the view. It's fun for some reason to open my mouth and stick out my tongue pretend I can feel it catch all the light that hits like little raindrops full of energy. 
\end{quote}



\subsection{Sense of physical embodiment and sensory awareness}

The students' reflections abound with details of a heightened somatic, sensory, and environmental awareness.

\begin{quote}
Suddenly, I felt so much more aware of the space around me. It was like my senses had truly awakened for the first time in a long time, and with it, my mind awakened, too. I was aware of everything from the tightness in the muscles on my face (which was unsettling, since I wasn't intentionally doing anything with it) to the feeling of the base of my spine connecting down through the dirt to be grounded with the earth. I was aware of everything from the sounds of the wind rustling through the trees to my own slow breath. I was aware of the scents of the different trees around me, the earthiness of the soil, the freshness of the forest. I was aware.  
\end{quote}


\begin{quote}
I felt as though my senses were hyper-aware of my surroundings, with every tree and flower bud being displayed in high definition. The bench I was sitting on was cool to the touch, but the sun tickled my bare shoulders with its warmth. As I absorbed the silence, I found myself noticing my environment much more so than before. I spent a solid five minutes watching a squirrel scamper up trees and chew noisily. I observed the ants marching along a crack in the cobblestones, steadfast in their pursuit of food. I watched the blossoms sway gently in the breeze, as if they were dancing. At no point did I feel the urge to take out my cell phone.  
\end{quote}


\begin{quote}
I walked around a bit too and felt all the different textures of various leaves and petals. I forgot what large variance there was in the feel of plants. One light purple flower had such soft petals that I just stood there rubbing the petals [thinking] about how gentle nature can be.  
\end{quote}




\subsection{Somatic relaxation and mental stillness}

A majority of students reported that the practices resulted in feelings of somatic relaxation and mental stillness. 

\begin{quote}
I closed my eyes.  I wanted to let \ldots \ nature take me away.  I let the breeze run across my face and the sun's rays find my skin.  There was a calming effect through my whole [being].  My mind went blank and my body went limp.  
\end{quote}

\begin{quote}
I closed my eyes and took a deep breath. The air was light, crisp and slightly cold. I took another deep breath and let my mind go. I sat in silence for a while in a state of acceptance and let go of my thoughts, my worries, and for a bit, the tasks at hand. I reminded myself to take slow, even, and steady breaths. I could feel my anxiety reduced, and my mind started focusing internally.  
\end{quote}

\begin{quote}
I started to notice how my body was responding to this moment of peacefulness. My muscles were loosened. My black hair had acquired pleasant warmth under the sun. Although I was relaxed, I didn't feel sleepy. In fact, I felt more alert than usual, more perceptive of my surroundings.  
\end{quote}
Many pointed out that the contemplative practices offered a dramatic contrast to their typical experience as college students.




\begin{quote}
Yeah, the stillness is weird. It's certainly very different from my everyday. And it's odd that I like it. I spend so much of the day doing anything I can not to be bored, but this stillness and complacency are oddly refreshing. Perhaps I shouldn't try so hard to fill every minute of my day with entertainment \ldots . It seems I design my day such that every moment is occupied, whether that design is a conscious decision or not. But yeah, maybe I should allow more for this. It's relaxing. It feels nice. 
\end{quote}


\begin{quote}
It is weird seeing everything so still \ldots . Compared to my usual day's commotion, it is a nice change of pace. I feel like I can really think and sort through my thoughts. I also notice that my body is almost as still as my surroundings. Especially for me, this is highly out of the ordinary---I am usually always running around or doing something---I think my body is thankful for the peace and calm.  
\end{quote}



\begin{quote}
It was liberating to not think about all the things I have to do all the time. It was nice to just take some time to take care of my brain and just listen to it and listen to my body \ldots . After doing this for a bit, I started thinking about the environment and my senses interacting with it \ldots . It was a pleasant experience.  
\end{quote}



\begin{quote}
Our lives as students \ldots \ are packed so full of constant stimulation, both positive and negative, that demands our attention that it leaves little time for personal reflection and peace unless you seek it out (something that I've never really done before). Even before I began to observe phenomena around me, this was a pretty powerful realization and actually helped me with some of the stress I've been feeling this quarter.  
\end{quote}


\subsection{Recognition of role of observation in science}

A number of students discussed how the practices allowed them to think about the central role of observation in the scientific endeavor.


\begin{quote}
I think that the origin of the study of science is human observation leading us to question and reflect on the physical world around us. By thinking about what we see in our everyday lives and wondering how or why it happens and then questioning why something else doesn't occur instead can lead us to struggle with the science we do know. This directly impacts us and makes us uncomfortable and that leads us to dig deeper, and discover more, therefore advancing our knowledge and our understanding of the world around us. When something we see cannot be explained by our science we know, we often need to look further to re-explain the physical happenings around us. I believe it is human nature to want to understand what we see around us, so as long as we take the time to look, science will always advance naturally.  
\end{quote}

\begin{quote}
[A contemplation] such as this one practices the idea of non­judgemental observation, something that I imagine is crucial in \ldots \ physics. For example, someone doing research may disregard or try to correct out some data that is unexpected in their experiments if they don't have an open mind. But if they do have an open mind, they may be intrigued by this fluctuation and actually alter the experiment to magnify this and could possibly discover the world's next big physics thing. 
\end{quote}



\begin{quote}
Physics [consists of] laws and theories about how the world behaves, so these \ldots \ have to come out of careful observations of our surroundings.  It is easy to forget this in a physics class, where we mostly just rely on conclusions that other physicists came to.  If we stop observing, none of these theories will ever be challenged and no new theories will ever be created.  
\end{quote}

\subsection{Curiosity in response to observations}

Many students described feeling a naturally emergent sense of curiosity in response to their observations, including in their reflections long lists of questions that arose while carrying out the practices.

\begin{quote}
This causes me to raise so many questions \ldots \ as I sit and witness nature.  Why do some of the tree branches remain in place and not bend towards the ground?  Why do some of the tree branches grow and bend towards the ground but never fall?  How strong/weak is this gravitational pull and how does the tree overcome it and grow so tall?  How do tiny insects like bees and flies overcome this pull and fly, yet with each jump we make as humans we are quickly drawn back to the ground?  
\end{quote}

\begin{quote}
As I walked to my computer science class [after finishing the contemplative practice], I noticed that everyone was looking at his/her cellphone. I wondered if they knew how the battery inside their cell phones kept a difference in voltage that mobilizes electrons, thus breathing life (electricity) to the apparatus. It is so easy nowadays to use electronic devices without having any idea about how they work. How do earphones transform electric signals into sounds? Why are some better than others? How do touchscreens work? How do credit cards work? How do they communicate to banks that a new transfer has been made? How do monitors work? I felt tremendously ignorant, but when I looked at my phone and saw the Google Chrome icon, I realized that the same technology that I did not fully understand could give me answers to my questions. It's so beautiful! We have answers at the tips of our fingers. And yet, in the most tragic manner, we ask less and less questions.  
\end{quote}

\begin{quote}
I try to focus on things that aren't quite apparent to me, and I notice that when I stare into the distance the wind frequently blows my hair across my eyes.  However, I don't notice the fluctuations of my hair interrupting my vision unless I think about it.  This makes me think about how my eyes work as complex lenses and how they \ldots \ form images that ignore the pieces of hair that consistently obstruct their view for fractions of a second.  Does it have something to do with the focal length being at a distance rather than right in front of my eye?  Possibly causing these images to be very out of focus and not easily distinguished in the sharp images created before me?  Or is it simply a function of our brains choosing to ignore images that my eyes are not directly focusing on?  
\end{quote}

\clearpage

\noindent Many students discussed the fact that curiosity lies at the foundation of the scientific endeavor.
\begin{quote}
I think [this practice] inspired questions in me and led me to think about the answers for the rest of the day. It was nice to think about these questions with no fear of getting the wrong answer. Speculating was fun and enriching. I think this type of speculation is what leads to amazing breakthroughs in physics and all other sciences because it is a way to provide answers that may be a little out of the box.  
\end{quote}


\begin{quote}
I think that contemplative practices such as this one \ldots \ can lead to discovery through observation. As I think about all the advances that we have, I cannot help but think that they arose because great scientists took more time to think and contemplate than we may do nowadays in the constant hustle and bustle of life. As someone who hopes to make a great discovery to help the world one day, I think taking time and reflecting is important for self-growth and discovery. I appreciated the opportunity this practice gave me in trying out this reflection and observation of the world.  
\end{quote}



\begin{quote}
I personally think that this practice of encouraging wild trains of thought is crucial to studying science, and particularly crucial to advancing science. Every experiment done has a purpose, and that purpose is attempting to answer a question. If no one takes the time to think of questions, how will we continue to expand experimentation? \ldots Curiosity is both the origin and the future of physics, and taking time to just contemplate the world around you is the first step in advancing science.  
\end{quote}
Several students described reconnecting with an innate curiosity and inquisitiveness that they last remembered experiencing as children.

\begin{quote}
When I stopped trying to forcefully observe, I actually began to notice things: the thoughts in my mind, the world surrounding me, and the greater questions beyond myself. Releasing myself of the fetters of society and individual limitations, I was able to more openly explore that singular moment, and its relation to the past, present, and future. The further I dissociated my thoughts from my preoccupations, the deeper my questions became about simpler ideas. Rather than encasing myself in what I have always known, I found myself thinking with childhood wonder about things that I have grown calloused to: the color of the sunset, the size of a rock, and the stars in the evening sky.  
\end{quote}

\begin{quote}
The glass windows of [the building] glitter in the sun, I look away. I think about light, light as a wave. It undergoes diffraction, and reflection, and refraction. I can see myself in a mirror because of light; rainbows happen because of light. I'm asking myself why I don't spend enough time outside, why I don't allow myself to take the time and appreciate my surroundings. [This institution] is beautiful. Nature is beautiful. I feel a bit like a child again.  
\end{quote}

\begin{quote}
As a child, [I used to] be fascinated with [my] surroundings, but nowadays I have found that fascination to have disappeared.  I have ceased wondering why something is the way it is and I find that disappointing.  How have I lost that drive?  How have I stopped taking the time to take in my surroundings and question why things are the way they are?  I'm striving to become a scientist and yet this is the first time I can recall questioning [physical] phenomena in a long time.  
\end{quote}

\begin{quote}
This practice really got me to observe [physical phenomena] at a deeper level than I typically do.  Not only that, but it let my mind wander and be creative.  This creativity used to happen all the time [when I was] a kid but going through college learning more and more specific things kind of kills \ldots \ any time that can be devoted to feeding that curiosity and creativity. 
\end{quote}
Inspired by their observations, students used this emergent sense of curiosity to experiment with centers of mass, damped oscillations, magnets, standing waves, and refractive indices, which they recorded in their videos.

\subsection{Intrinsic motivation to study physics}

Along with a sense of curiosity, students also experienced a sense of intrinsic motivation to study physics in response to the desire to find explanations for physical phenomena.


\begin{quote}
I think contemplative practices \ldots \ are important in physics because they give you time to think without necessarily looking for an answer or able to even compute an answer. It is specifically this type of meandering thought that I think is very lacking in physics classes, where we are expected to be focused on solving a specific problem in a specific amount of time. Thinking about physics in nature is also a reminder that physics is an attempt to understand the rules that govern the formation of our universe. It's a nice reminder of the applicability and motivation behind studying physics, and how it all ties back to the natural world. [This type of practice] also gives you time to consider your own relationship to physics, and your place in the universe as a sentient being that has the opportunity to understand it in some way.
\end{quote}

\begin{quote}
It is important to complement our problem solving skills with an appreciation of why the problems matter and what they explain.  The simplest natural processes are dictated by the laws we have been learning throughout the year.  Engaging with our physical world not only encourages us to appreciate what we have learned, but also encourages us to find answers for phenomena we may not understand.  
\end{quote}


\begin{quote}
I do not often stop to think about how important some of the topics I have learned up to this point are, but this exercise really made me think about it. It makes me more motivated to think about what I am learning in class and in my textbooks and stop more often to think about why I am learning it, and why it is important. How can these principles be applied in my everyday life; how do they affect the way I interact with others and my surroundings?  
\end{quote}


\begin{quote}
This assignment really touched me, as cheesy as that sounds \ldots . I have always been drawn to and loved everything science. That being said, it has lost a bit of its magic as I've gotten older because it became associated with equations and theories and confusing concepts that I had to memorize for a good grade. It was still interesting to me, but in a more stressful way \ldots . I really appreciate that these assignments are offered to us \ldots \ because [they bring] back some of the awe and magic of how amazing science can be.  
\end{quote}


\subsection{Awareness of cognitive and emotional processes}

In addition to the specific cognitive insights discussed above, the students also reported an overall increased awareness of their cognitive and emotional processes.

\begin{quote}
As a human I will always have that natural tendency to be able to perform physical actions without a second thought, such as catching a ball, playing an instrument, or driving a car. But the other half of my brain that I cultivate within class is more logical. It needs instructions---explanations and equations for how things work---in order to comprehend things on more than just an instinctual level. There during that sunny afternoon, I realize that I must get to the point where I can understand physics at the intersection of these two exceedingly different approaches. By observing the world around me and seeing first-hand how it relates to all the numbers and variables I manipulate on paper every day, I am one step closer to reaching this intersection. I haven't gotten there yet. But every time I go up to the observatory, I know I'll be able to escape the world of textbooks and grades. I'll get to see physics in its most natural, human state. And I'll regain confidence that someday, I'll have my two sides of my brain working in tandem as a physicist.  
\end{quote}


\begin{quote}
With the silence, there is stillness. This makes sense, because moving objects are also often the producers of noise. But there is also a mental stillness inside me \ldots . The silence and the stillness, they force me to notice my emotions and recognize their causes. I feel what I had been avoiding, that I am tired and a little scared for finals, but I also see that I am happy in a way that feels much more stable than I have known happiness to be in the past few years.  
\end{quote}

\begin{quote}
It was cold outside during my walk \ldots, and my extremities are still numb. The room is much warmer, and I visualize the transfer of heat from the air into my body, into my blood that is carried all the way into every corner of me. I am slowly and gently thawing. It is a soothing thought. My head is clear and sharp, agile. My mind feels boundless, as if it is expanding to fill and feel this room.  
\end{quote}

\begin{quote}
In this moment, I feel calmer. I feel like I have all the time in the world. I feel less stressed out, less anxious. I feel more in touch with how I feel, both emotionally and physically.  
\end{quote}

\subsection{Skepticism}

Out of all of the students who carried out these practices, only one openly expressed doubt that there was any value in them.
\begin{quote}
Overall, I'm not really sure that this experience really helped deepen my connection to physics or science or the world or anything like that \ldots . Sitting down [and] taking a moment to clear your head can certainly be relaxing, but it hasn't really led to me having any deep insights regarding physics, or anything else in life, really.  
\end{quote}
In a handful of cases, students described feeling skeptical prior to doing the practices, but then realizing that there was something to be learned from them.

\begin{quote}
I honestly thought that these contemplative practices would be less interesting, but I am very happy to say that they have been a wonderful experience. They reminded me of the reasons why I love science, and that I should not stop asking questions about my surroundings. And of course, they helped me see the many applications of the physics concepts that we learn in the classroom in a more personal way. The fact that all I needed to do was to relax and truly observe and appreciate nature made the experience very enjoyable.  
\end{quote}


\begin{quote}
I suppose I'm done now. This was fun; I'm glad you made me do it. I thought it was going to be pretty much a waste of time but it didn't take that long and I enjoyed it. Hopefully I answered the prompts in a satisfactory way.  
\end{quote}


\begin{quote}
Wow I have to admit that was a pretty cool experience. After reading the introduction and procedure I thought that this was going to be tedious but then I reminded myself to just let go and try it out. And I must say that I had fun doing it. I like questioning the world around me as well as questioning the status quo.  
\end{quote}




\section{Conclusion} 
\label{conclusion}

Spanning three academic years and three institutions of higher education, this work represents a qualitative, exploratory analysis of the impact of contemplative practices on the learning experience in the undergraduate physics curriculum.  Because it was the first work of its kind, we were committed to letting the students' experiences, as expressed through their written reflections, speak largely for themselves.  In future studies, we hope to administer the CLASS at the beginning and end of the course in order to collect the type of quantitative data that would allow us to compare contemplative practices with other interventions aimed at integrating formal theory and personal experience in the undergraduate physics curriculum.  Furthermore, administration of instruments such as the Mindful Attention Awareness Scale (MAAS) and the Five Facet Mindfulness Questionnaire (FFMQ) would allow us to assess the general effects of these contemplative practices on students' awareness of their present, embodied experience \cite{Brown2003,Baer2008}.

The contemplative practices presented in this work invite students in the undergraduate physics curriculum to integrate formal theory with their personal, embodied experience.  In the process of directing their attention into their bodies, expanding their sensory awareness, and beholding the physical phenomena occurring in their surroundings, however, students discover far more than manifestations of formal physics principles.  By engaging with these practices, students found relaxation, mental stillness, and a heightened sense of physical embodiment, in great contrast to a college experience typically characterized by stress, anxiety, and neverending activity.  They also experienced insights about the essential place of observation in the scientific endeavor and the spontaneous manifestation of curiosity in response to such observations.  Many students reported having felt disconnected from the innate curiosity that originally inspired them to study science, and emerged from the practices with a renewed sense of motivation.  Finally, students described experiencing a heightened awareness of their cognitive and affective processes, in alignment with one of the original, yet often forgotten, goals of a university education: the development of a capacity for deep self-awareness and reflection.  

One of the great paradoxes of contemporary science is that in spite of its sophistication, human beings feel increasingly disconnected from the very realm it so successfully and powerfully describes. We are educated to believe that the true nature of physical reality is not as we experience it, leading us to question the applicability of our immediate perceptions to the abstractions of the scientific canon.  At the end of the eighteenth century, following the success of the quantitative methods of Galileo, Descartes, Newton, and others, Johann Wolfgang von Goethe wrote, in a passionate call for the relevance of our qualitative and deeply personal experience of the natural world,
\begin{quote}
None of the human faculties should be excluded from scientific activity.  The depths of intuition, a sure awareness of the present, mathematical profundity, physical exactitude, the heights of reason and sharpness of intellect together with a versatile and ardent imagination, and a loving delight in the world of the senses---they are all essential for a lively and productive apprehension of the moment \cite{Naydler1996}.
\end{quote}


\addcontentsline{toc}{section}{Acknowledgements}
\section*{Acknowledgements}

The authors would like to express their profound gratitude to the many students whose deep open-mindedness and open-heartedness made this work possible.  We would also like to thank Elam Coalson and Calais Larson for their thoughtful contributions and constructive feedback.  Our research was supported by grants from the Carolyn Grant `36 Endowment, the URSI Program, and the Ford Scholars Program at Vassar College, and by the Helmsley Charitable Trust at Yale University.  

\clearpage

\addcontentsline{toc}{section}{References}
\bibliographystyle{unsrt}
\bibliography{standard-bibliography}

\begin{thebibliography}{10}

\bibitem{Redish2003b}
Edward~F. Redish.
\newblock {\em {Teaching Physics with the Physics Suite}}.
\newblock John Wiley {\&} Sons, Hoboken, NJ, 2003.

\bibitem{NRC2013}
{National Research Council Committee on Undergraduate Physics Education
  Research and Implementation}.
\newblock {\em {Adapting to a Changing World: Challenges and Opportunities in
  Undergraduate Physics Education}}.
\newblock The National Academies Press, Washington, DC, 2013.

\bibitem{Redish1998}
Edward~F. Redish, Jeffery~M. Saul, and Richard~N. Steinberg.
\newblock {Student expectations in introductory physics}.
\newblock {\em American Journal of Physics}, 66(3):212--224, 1998.

\bibitem{Adams2006}
Wendy~K. Adams, Katherine~K. Perkins, Noah~S. Podolefsky, Michael Dubson,
  Noah~D. Finkelstein, and Carl~E. Wieman.
\newblock {New instrument for measuring student beliefs about physics and
  learning physics: The Colorado Learning Attitudes about Science Survey}.
\newblock {\em Physical Review Special Topics - Physics Education Research},
  2(1):1--14, 2006.

\bibitem{Madsen2015}
Adrian Madsen, Sarah~B. McKagan, and Eleanor~C. Sayre.
\newblock {How physics instruction impacts students' beliefs about learning
  physics}.
\newblock {\em Physical Review Special Topics - Physics Education Research},
  11(1):010115, 2015.

\bibitem{Crouch2001}
Catherine~H. Crouch and Eric Mazur.
\newblock {Peer Instruction: Ten years of experience and results}.
\newblock {\em American Journal of Physics Teachers}, 69(9):970--977, 2001.

\bibitem{May2002}
David~B. May and Eugenia Etkina.
\newblock {College physics students' epistemological self-reflection and its
  relationship to conceptual learning}.
\newblock {\em American Journal of Physics}, 70(12):1249--1258, 2002.

\bibitem{Wieman2005a}
Carl~E. Wieman and Kathy~K. Perkins.
\newblock {Transforming physics education}.
\newblock {\em Physics Today}, 58(11):36--41, 2005.

\bibitem{Chu2007}
Hye-Eun Chu, David~F. Treagust, and A.~L. Chandrasegaran.
\newblock {Na{\"{i}}ve students' conceptual development and beliefs: The need
  for multiple analyses to determine what contributes to student success in a
  university introductory physics course}.
\newblock {\em Research in Science Education}, 38:111--125, 2008.

\bibitem{Crouch2014}
Catherine~H. Crouch and Kenneth Heller.
\newblock {Introductory physics in biological context: An approach to improve
  introductory physics for life science students}.
\newblock {\em American Journal of Physics}, 82(5):378--386, 2014.

\bibitem{Lasry2016}
Nathaniel Lasry, Elizabeth Charles, and Chris Whittaker.
\newblock {Effective variations of peer instruction: The effects of peer
  discussions, committing to an answer, and reaching a consensus}.
\newblock {\em American Journal of Physics}, 84(8):639--645, 2016.

\bibitem{Webb2017}
D.~J. Webb.
\newblock {Concepts first: A course with improved educational outcomes and
  parity for underrepresented minority groups}.
\newblock {\em American Journal of Physics}, 85(8):628--632, 2017.

\bibitem{Barbezat2014}
Daniel~P. Barbezat and Mirabai Bush.
\newblock {\em {Contemplative Practices in Higher Education}}.
\newblock Jossey-Bass, San Francisco, CA, 2014.

\bibitem{Kabat-Zinn2006}
Jon Kabat-Zinn.
\newblock {Mindfulness-based interventions in context: Past, present, and
  future}.
\newblock {\em Clinical Psychology: Science and Practice}, 10(2):144--156,
  2006.

\bibitem{Ospina2007}
Maria~B. Ospina, Kenneth Bond, Mohammad Karkhaneh, Lisa Tjosvold, Ben
  Vandermeer, Yuanyuan Liang, Liza Bialy, Nicola Hooton, Nina Buscemi, Donna~M.
  Dryden, and Terry~P. Klassen.
\newblock {Meditation practices for health: State of the research}.
\newblock {\em Evidence Report/Technology Assessment}, (155):1--263, 2007.

\bibitem{Bohlmeijer2010}
Ernst Bohlmeijer, Rilana Prenger, Erik Taal, and Pim Cuijpers.
\newblock {The effects of mindfulness-based stress reduction therapy on mental
  health of adults with a chronic medical disease: A meta-analysis}.
\newblock {\em Journal of Psychosomatic Research}, 68(6):539--544, 2010.

\bibitem{Hayes2011}
Steven~C. Hayes, Matthieu Villatte, Michael Levin, and Mikaela Hildebrandt.
\newblock {Open, aware, and active: Contextual approaches as an emerging trend
  in the behavioral and cognitive therapies}.
\newblock {\em Annual Review of Clinical Psychology}, 7(1):141--168, 2011.

\bibitem{Kerr2013}
Catherine~E. Kerr, Matthew~D. Sacchet, Sara~W. Lazar, Christopher~I. Moore, and
  Stephanie~R. Jones.
\newblock {Mindfulness starts with the body: Somatosensory attention and
  top-down modulation of cortical alpha rhythms in mindfulness meditation}.
\newblock {\em Frontiers in Human Neuroscience}, 7:1--15, 2013.

\bibitem{Buchholz2015}
Laura Buchholz.
\newblock {Exploring the promise of mindfulness as medicine}.
\newblock {\em Journal of the American Medical Association},
  314(13):1327--1329, 2015.

\bibitem{Jha2007}
Amishi~P. Jha, Jason Krompinger, and Michael~J. Baime.
\newblock {Mindfulness training modifies subsystems of attention}.
\newblock {\em Cognitive, Affective and Behavioral Neuroscience},
  7(2):109--119, 2007.

\bibitem{Shapiro2008}
Shauna~L. Shapiro, Kirk~Warren Brown, and John~A. Astin.
\newblock {Toward the integration of meditation into higher education: A review
  of research}.
\newblock {\em Teachers College Record}, 113(3):493--528, 2008.

\bibitem{Morrison2014}
Alexandra~B. Morrison, Merissa Goolsarran, Scott~L. Rogers, and Amishi~P. Jha.
\newblock {Taming a wandering attention: Short-form mindfulness training in
  student cohorts}.
\newblock {\em Frontiers in Human Neuroscience}, 7:1--12, 2014.

\bibitem{Rechtschaffen2014}
Daniel Rechtschaffen.
\newblock {\em {The Way of Mindful Education: Cultivating Well-Being in
  Teachers and Students}}.
\newblock W. W. Norton and Company, New York, NY, 2014.

\bibitem{Short2015}
Megan~M. Short, Dwight Mazmanian, Lana~J. Ozen, and Michel B{\'{e}}dard.
\newblock {Four days of mindfulness meditation training for graduate students:
  A pilot study examining effects on mindfulness, self-regulation, and
  executive function}.
\newblock {\em The Journal of Contemplative Inquiry}, 2(1):37--48, 2015.

\bibitem{Palmer2017}
Parker Palmer.
\newblock {\em {The Courage to Teach: Exploring the Inner Landscape of a
  Teacher's Life}}.
\newblock John Wiley {\&} Sons, Hoboken, NJ, 2017.

\bibitem{Hart2004}
Tobin Hart.
\newblock {Opening the contemplative mind in the classroom}.
\newblock {\em Journal of Transformative Education}, 2(1):28--46, 2004.

\bibitem{Zajonc2006a}
Arthur Zajonc.
\newblock {Contemplative and transformative pedagogy}.
\newblock {\em Kosmos Journal}, 5(1):1--3, 2006.

\bibitem{Bush2011}
Mirabai Bush.
\newblock {Mindfulness in higher education}.
\newblock {\em Contemporary Buddhism}, 12(1):183--197, 2011.

\bibitem{Levy2016}
David Levy.
\newblock {\em {Mindful Tech: How to Bring Balance to Our Digital Lives}}.
\newblock Yale University Press, New Haven, CT, 2016.

\bibitem{Schneiderman2013}
Jill Schneiderman.
\newblock {Ground truth: Investigations of Earth simultaneously spiritual and
  scientific}.
\newblock In Jing Lin, Rebecca Oxford, and Edward Brantmeier, editors, {\em
  Re-Envisioning Higher Education: Embodied Pathways to Wisdom and Social
  Transformation}. Information Age Publishing, Charlotte, NC, 2013.

\bibitem{Zajonc2013}
Arthur Zajonc.
\newblock {Contemplative pedagogy: A quiet revolution in higher education}.
\newblock {\em New Directions for Teaching and Learning}, (134):83--94, 2013.

\bibitem{Francl2016}
Michelle Francl.
\newblock {Practically impractical: Contemplative practices in science}.
\newblock {\em Journal of Contemplative Inquiry}, 3(1):21--34, 2016.

\bibitem{Ray2008}
Reginald~A. Ray.
\newblock {\em {Touching Enlightenment: Finding Realization in the Body}}.
\newblock Sounds True, Boulder, CO, 2008.

\bibitem{Ray2016}
Reginald~A. Ray.
\newblock {\em {The Awakening Body: Meditation for Discovering our Deepest
  Life}}.
\newblock Shambhala Publications, Boston, MA, 2016.

\bibitem{Rome2014}
David~I. Rome.
\newblock {\em {Your Body Knows the Answer}}.
\newblock Shambhala Publications, Boston, MA, 2014.

\bibitem{Trungpa2008}
Ch{\"{o}}gyam Trungpa.
\newblock {\em {True Perception: The Path of Dharma Art}}.
\newblock Shambhala Publications, Boston, MA, 2008.

\bibitem{Karr2011}
Andy Karr and Michael Wood.
\newblock {\em {The Practice of Contemplative Photography: Seeing the World
  with Fresh Eyes}}.
\newblock Shambhala Publications, Boston, MA, 2011.

\bibitem{Brown2003}
Kirk~Warren Brown and Richard~M. Ryan.
\newblock {The benefits of being present: Mindfulness and its role in
  psychological well-being}.
\newblock {\em Journal of Personality and Social Psychology}, 84(4):822--848,
  2003.

\bibitem{Baer2008}
Ruth~A. Baer, Gregory~T. Smith, Emily Lykins, Daniel Button, Jennifer
  Krietemeyer, Shannon Sauer, Erin Walsh, Danielle Duggan, and J.~Mark~G.
  Williams.
\newblock {Construct validity of the five facet mindfulness questionnaire in
  meditating and nonmeditating samples}.
\newblock {\em Assessment}, 15(3):329--342, 2008.

\bibitem{Naydler1996}
Jeremy Naydler.
\newblock {\em {Goethe on Science: An Anthology of Goethe's Scientific
  Writings}}.
\newblock Floris Books, Edinburgh, UK, 1996.

\end{thebibliography}

\clearpage

\appendix
\appendixpage

\section{Guidelines for the sensory meditation}

\subsection{Procedure}

Individually, and in silence, take a seat somewhere on the university campus, preferably away from particular spaces and pathways that tend to inundate with people.  Look at the space around you.  Notice what is particular about the spot you have chosen---what can you see, hear, smell, touch?  Begin to take in the stillness of the space.  Finally, when you feel settled, observe the physical phenomena taking place all around you.  What do you notice?  What effect are these phenomena having on your mind?  Please carry out this practice on at least a couple of occasions. 

\subsection{Reflection}

Please write up a reflections on this practice, addressing the following questions: 
\begin{enumerate}
\item How did you experience this practice? 
\item Do you feel that practices such as this one can inform and/or complement the formal study of physics? 
\end{enumerate}
Your reflection should be approximately one to two pages in length.

\clearpage

\section{Guidelines for the contemplative videography}

\subsection{Procedure}

Select one physical law, principle, or concept from the introductory physics curriculum.  Using a video recording device (typically, your smart phone), spend a week filming objects, systems, and/or interactions in your physical surroundings that illustrate and illuminate the concept you have chosen.  Compile your clips into a short film.  If at all, use special effects sparingly and thoughtfully.  

\subsection{Reflection}

Please write up a reflections on this practice, addressing the following questions: 
\begin{enumerate}
\item How did you experience this practice? 
\item Do you feel that practices such as this one can inform and/or complement the formal study of physics? 
\end{enumerate}
Your reflection should be approximately one to two pages in length.

\end{document}